\documentstyle[PASJadd,epsf]{PASJ95}

\begin{document}

\title{Conversion Law of Infrared Luminosity to Star Formation
       Rate for Galaxies}

\author{Akio K. {\sc Inoue},\ Hiroyuki {\sc Hirashita}\thanks{Research
        Fellow of Japan Society for the Promotion of Science}, 
        and Hideyuki {\sc Kamaya}
\\[12pt]
{\it Department of Astronomy, Faculty of Science, Kyoto University,
Sakyo-ku, Kyoto 606-8502}\\
{\it E-mail(AKI): inoue@kusastro.kyoto-u.ac.jp}}

\abst{
We construct a new algorithm for estimating the star formation rate (SFR)
of galaxies from their infrared (IR) luminosity by developing the theory
of the IR emission from a dusty H II region.
The derived formula is
${\rm SFR}/(\MO\, {\rm yr^{-1}})=\{3.3\times 10^{-10}(1-\eta)/
       (0.4-0.2f+0.6\,\epsilon)\}(L{\rm ^{obs}_{IR}}/\LO)$,
where $f$ is the fraction of ionizing photons absorbed by hydrogen,
$\epsilon$ is  the efficiency of dust absorption for nonionizing
photons from OB stars, and $\eta$ is the cirrus fraction of observed IR
luminosity.
The previous conversion formulae of SFR from the IR luminosity is
applicable to only the case where the observed IR luminosity is nearly
equal to the bolometric luminosity (starburst galaxies etc.), except for 
some empirical formulae.
On the other hand, our theoretical SFR is applicable to galaxies even
with a moderate star formation activity.
That is, our simple and convenient formula is significantly useful for
estimating the SFR of various morphologies and types of galaxies --- from
early elliptical to late spiral and irregular galaxies, or from active
starburst to quiescent galaxies --- as far as they have neither an abnormal
dust-to-gas ratio nor an evident active galactic nucleus.
}

\kword{galaxies: fundamental parameters --- infrared: galaxies --- ISM:
dust --- stars: formation --- methods: analytical}

\maketitle
\thispagestyle{headings}

\section{Introduction}

There are a number of studies estimating the star formation rate (SFR)
of galaxies by using various observational quantities.
Indeed, we can estimate the SFR from colors of galaxies, hydrogen
recombination lines, ultra-violet (UV) continuum and so on (Kennicutt 1998a).
Of course, infrared (IR) radiation is also a very important tracer of
the SFR.
This is because some fraction of ionizing and noninonizing photons
from star forming regions are absorbed by dust grains which
exist in or nearby these regions, and their energy is reradiated in IR
spectral range (Soifer et al. 1987).

In fact, most active star forming galaxies emit almost all of their
radiative energy in IR (Soifer et al. 1987).
For such galaxies, we can assume that their observed IR luminosity is
nearly equal to their bolometric luminosity. 
By adopting this assumption and a proper population synthesis model,
Kennicutt (1998b) and Thronson \& Telesco (1986) have estimated SFRs
from the IR luminosity for starburst galaxies and for active dwarf
galaxies, respectively.
Likewise, taking into account the energy balance of such active star
forming galaxies, we can also calculate their SFR from the IR
luminosity (Devereux \& Young 1991).

For normal galaxies with a moderate star formation activity, however, the
situation is more complex, because we can no longer assume their IR
luminosity to be their bolometric one.
In such cases, only empirical methods have been made to date.
For example, Buat \& Xu (1996) determined the SFR in terms of the far-IR
luminosity, calculating the ratio of it to the UV luminosity.
However, their SFR is valid for late (Sb--) spiral galaxies only, since
the ratio of far-IR to UV luminosity for early spiral
(Sa--Sab) galaxies systematically differs from that for late spiral
galaxies.
Thus, we will establish more general method for the estimation of the
SFR from the IR luminosity in this Letter.

In general, we consider that massive young stars are formed
in H II regions.
In many H II regions, IR radiation is detected and it has a good
correlation with thermal free-free radio emission, which comes directly
from the ionized regions and is directly related to massive star
formation (e.g., Spitzer 1978).
Therefore, dust grains exist in or nearby H II
regions, and they absorb a part of radiation from young massive stars
and reradiate this absorbed energy in IR wavelengths. 
Petrosian et al. (1972) estimated the IR luminosity radiated from dusty
H II regions with a simple analytic approximation.
Our analysis is based on their result.

Unfortunately, not all the IR radiation originates from star forming regions
in galaxies.
The observed IR radiation from dust is the sum of the ``warm'' component 
distributed in or nearby star forming regions and the ``cool'' 
component (called cirrus component) distributed diffusely in
the interstellar medium far from such regions (e.g., Helou 1986).
Thus, the SFR calculated from the observed {\it raw} IR luminosity is
clearly overestimated.
When we estimate the SFR from IR luminosity, thus, the cirrus
component must be subtracted.
In this Letter, we will subtract the cirrus component according to
Lonsdale-Persson \& Helou (1987), which presented a model of the cirrus
fraction by using the ratio of IRAS 60$\mu$m and 100$\mu$m fluxes.
Also, we can estimate this fraction by applying a proper radiative
transfer model for reproducing multi-wavelength data of galaxies (e.g.,
Silva et al. 1998, Efstathiou et al. 2000).
However, the fraction of the cirrus component remains still uncertain.

We stress again that only empirical studies on SFR from the IR luminosity of
galaxies with a moderate star formation activity have been made to date.
Thus, we develop the result of Petrosian et al. (1972) and construct 
a new algorithm to derive the SFR from the IR luminosity in this Letter.
Thereby, we obtain a convenient formula, and calculate the SFR of
galaxies from their IR luminosity.
This Letter contains the following sections: we review the IR
luminosity of a dusty H II region in \S 2, our formula of SFR is
constructed in \S 3, we discuss some results derived from our
formula in \S 4, and the conclusions of this Letter are summarized in
the last section.

\section{Infrared Luminosity of H II regions}

First of all, let us look closely at the wavelength range of IR radiation 
from H II regions.
We consider IR radiation to be mainly thermal black body radiation of dust
heated by young massive stars in such regions.
In fact, IR spectral energy distribution (SED) observed in H II regions
can be fitted by the black body radiation of about 30--50 K (Soifer et
al. 1987).
By the way, the temperature of dust within ionized regions is higher
than that of dust out of there.
That is, the peak wavelength of dust-IR radiation depends on the distance of
dust grains from ionizing stars.
It is, however, complex to take account of more than two temperatures of
dust and this is beyond the scope of this Letter.
Instead of determining exact SED, we consider the dust luminosity in the
whole range of 8--1000~$\mu{\rm m}$, which covers almost all the
emission from dust.
The increment owing to stars at the shorter wavelength in this range is
assumed to be subtracted.

Next, we see the result of Petrosian et al. (1972).
When the Case B approximation, which is an assumption of the large
optical depth for every Lyman-emission-line photon (e.g., Osterbrock 1989),
can be applied to a dusty H II region, Petrosian et al. (1972) derived
the following equation for the dust-IR luminosity, $L^{\rm dust}_{\rm
IR}(8-1000\mu{\rm m})$,   
\begin{eqnarray}
 &L^{\rm dust}_{\rm IR}{\rm (8-1000\mu m)}
 =L({\rm Ly\alpha})+(1-f){\langle h\nu\rangle}_{\rm ion}S & \nonumber \\
 & + L_{\rm nonion}(1-e^{-\tau}).&
 \label{pet}
\end{eqnarray}
Here, $L({\rm Ly\alpha})$ is the luminosity of Lyman-$\alpha$ emission
line, $f$ is the fraction of ionizing photons absorbed by hydrogen,
$\langle h\nu\rangle_{\rm ion}$ is the averaged energy of ionizing
photons, $S$ is the number of ionizing photons from central sources per
unit time, $L_{\rm nonion}$ is the luminosity of nonionizing photons, and
$\tau$ is the optical depth of dust for nonionizing photons.

They assumed that all Lyman-$\alpha$ photons produced by hydrogen
recombination processes in a H II region are absorbed by dust within
this region and reemitted as IR radiation because of their large optical
depth (i.e. Case B).
This is reflected in the first term of the right hand side of equation
(\ref{pet}).
The second term represents the energy of ionizing photons absorbed
directly by dust within the ionized region.
Then, the last term denotes the energy of nonionizing radiation absorbed 
by dust in or nearby ionized region, especially, in molecular clouds
surrounding this region.

We now consider the luminosity of the Lyman-$\alpha$ emission line within
H II regions.
Under Case B approximation, every ionizing photon will eventually form one
hydrogen atom with the $n=2$ level.
In this process, about two-thirds of recombining electrons will reach
the 2p state and go down to 1s, emitting a Lyman-$\alpha$ photon.
The remaining one-third of recombining electrons will reach the 2s state,
and two continuum photons will be emitted simultaneously within 1 second
because the transition from the 2s to the 1s is forbidden for any one
photon process (e.g., Spitzer 1978).
Thus, $L({\rm Ly\alpha})$ in terms of $S$ is $0.67h\nu_{\rm
Ly\alpha}fS$, where $\nu_{\rm Ly\alpha}$ is the frequency at the
Lyman-$\alpha$ emission.

The luminosity of ionizing photons from central sources,
$L_{\rm ion}$, is written as $\langle h\nu \rangle_{\rm ion}S$.
When we set $h\nu_{\rm Ly\alpha}$=10.2~eV and $\langle h\nu
\rangle_{\rm ion}\sim$ 15~eV, the luminosity of Lyman-$\alpha$ is given
by
\begin{equation}
 L({\rm Ly\alpha})
 =0.67\frac{h\nu_{\rm Ly\alpha}}{\langle h\nu \rangle_{\rm ion}}fL_{\rm ion}
 \simeq 0.45fL_{\rm ion}.
\end{equation}
Therefore, equation (\ref{pet}) is reduced to
\begin{equation}
 L^{\rm dust}_{\rm{IR}}{\rm (8-1000\mu m)}
 =(1-0.55f)L_{\rm ion}+\epsilon L_{\rm nonion},
 \label{IR}
\end{equation}
where $\epsilon \equiv 1-e^{-\tau}$ is an averaged
dust-absorption-efficiency of nonionizing photons from central sources
in H II regions.

\section{Derivation of Star Formation Rate}

We will derive a new convenient formula to determine the SFR
from the observed IR luminosity.
Our result is presented in equation (\ref{sfr}) in this section.
First, in order to estimate the recent SFR, we assume that all of the
total energy radiated from star forming regions is emitted by only young
massive stars with the spectral types of O and B on the main-sequence. 
We fit the mass-luminosity relation of main-sequence stars
from the table 3.13 in Binney \& Merrifield (1998) as
\begin{equation}
 \log l(m)= \cases{\log m +4 & for $m \ge 20 \MO$\cr
                    4 \log m & for $m < 20 \MO$},
 \label{masslumi}
\end{equation}
where $m$ is the stellar mass in solar unit and $l(m)$ is the stellar
luminosity in solar unit as a function of $m$.

We now must choose a initial mass function (IMF), $\psi(m)$, but the
choice of it may affect our calculation of the SFR.
There are two uncertainties in adopting a specific IMF: the slope of
$\psi(m)$, and the upper and lower cut-off of $m$.
In this Letter, we adopt the Salpeter's IMF, $\psi(m)\propto
(m/\MO)^{-2.35}$ (Salpeter 1955) and the mass range of 0.1 -- 100 $\MO$,
since they seem to be widely applicable.
We normalize the IMF by $\int \psi(m) dm = 1$.
We will discuss the uncertainty about the choice of an IMF more closely
in \S 4.

Next, we consider the mass range of O stars is 20 -- 100 $\MO$
and that of B stars is 3 -- 20 $\MO$.
Assuming that the total luminosity from star forming regions, $L{\rm
^{SF}_{total}}$, is the bolometric luminosity of O and B stars, $L{\rm
^{bol}_{OB}}$, we obtain
\begin{equation}
 L{\rm ^{SF}_{total}}=L{\rm ^{bol}_{OB}}
 =a \int^{100\MO}_{3\MO} l(m)\psi(m)dm,
 \label{lob}
\end{equation}
where $a$ is a normalization so that $M_*=a\int m\psi(m)dm$, when we
represent $M_*$ as the stellar mass formed newly in star forming regions. 
Since we adopt the Salpeter's IMF, the bolometric luminosity of O stars
and that of B stars are determined as $L{\rm
^{bol}_{Otype}}=0.8L{\rm ^{bol}_{OB}}$ and $L{\rm
^{bol}_{Btype}}=0.2L{\rm ^{bol}_{OB}}$, respectively.

Let us discuss what type of stars emits ionizing photons dominantly.
According to a model calculation by Panagia (1973), main-sequence stars
of B type hardly emit these photons whereas those of O type radiate
them.
However, not all photons from O type stars have shorter wavelength than
Lyman limit (912\AA).
The energy fraction of ionizing photons from O5 stars is 0.6 and that
of O9 stars is 0.2 (Panagia 1973). 
Then, if we adopt 60 $\MO$ and 20 $\MO$ as the mass of O5 and O9 stars,
respectively, and when the energy fraction of ionizing photons from
stars of mass $m$ is denoted by $\alpha(m)$, we get the following
relation approximately,
\begin{equation}
 \alpha(m/\MO) = 0.01(m/\MO).
 \label{ionfrac}
\end{equation}
Using equation (\ref{ionfrac}), we can calculate the ionizing
luminosity of main-sequence O type stars.
Then, we get $L{\rm ^{ion}_{Otype}}=0.5L{\rm ^{bol}_{Otype}}=0.4L{\rm
^{bol}_{OB}}$.
Now we assume that B type stars radiate nonionizing photons only.
That is,
\begin{eqnarray*}
 L_{\rm ion}&=&L{\rm ^{ion}_{Otype}}, \\
 L_{\rm nonion}&=&(L{\rm ^{bol}_{Otype}}-L{\rm ^{ion}_{Otype}})
                +L{\rm ^{bol}_{Btype}}.
\end{eqnarray*}
Thus, $L_{\rm ion}$ and $L_{\rm nonion}$ in equation (\ref{IR}) is
reduced to $0.4L{\rm^{bol}_{OB}}$ and $0.6L{\rm^{bol}_{OB}}$,
respectively.
We now obtain the following expression of the dust-IR luminosity
of star forming regions, $L^{\rm dust}_{\rm IR}(8-1000\mu{\rm m})$,
\begin{equation}
 L^{\rm dust}_{\rm IR}(8-1000\mu{\rm m})
 =(0.4-0.2f+0.6\,\epsilon)L{\rm ^{bol}_{OB}}.
\end{equation}

In general, the observed far-IR luminosity contains both the component
originated from star forming regions and that from the diffuse cirrus
(e.g., Helou 1986).
When we represent the cirrus fraction as $\eta$, the total
luminosity of star forming regions in a galaxy is, therefore, given by
\begin{equation}
 L{\rm ^{SF}_{total}}=L{\rm ^{bol}_{OB}}=
   \frac{1-\eta}{0.4-0.2f+0.6\,\epsilon}\,
   L{\rm ^{obs}_{IR}}({\rm 8-1000\mu m}), 
 \label{lob2}
\end{equation}
where $L{\rm ^{obs}_{IR}}(8-1000\mu{\rm m})$ is the observed dust-IR
luminosity of this galaxy.
Furthermore, since $M_*=a\int m\psi(m)dm$, the normalization $a$ is
removed by using equation (\ref{lob}):
\begin{equation}
 \frac{M_*}{\MO}=\frac{\int_{0.1\MO}^{100\MO}m\psi(m)dm}
  {\int_{3\MO}^{100\MO}l(m)\psi(m)dm}\,\frac{L{\rm ^{bol}_{OB}}}{\LO}
 = 1.1\times 10^{-3}\,\frac{L{\rm ^{bol}_{OB}}}{\LO}.
 \label{stellarmass}
\end{equation}

We now need to consider the time-scale of star formation because the SFR
is the mass $M_*$ divided by this time-scale if the SFR is assumed to be
constant during it.
Since we are interested in the recent SFR of galaxies, we regard it as the 
representative life time of main-sequence OB stars.
The stars on the main-sequence generate their radiative energy by the
nuclear hydrogen reaction.
Thus, the life-time of this phase is given by the following function of
mass of stars,
\begin{equation}
 t(m) = 0.0067\times\frac{\beta m c^2}{l(m)}\,,
\end{equation}
where $c$ is the light speed, $\beta$ denotes the fraction of mass
consumed during this phase and $l(m)$ is the mass-luminosity relation
given by equation~(\ref{masslumi}).
According to Schwarzchild (1958), the mass fraction, $\beta$, through
the CNO cycle is 0.13.
Therefore, we obtain the life time of main-sequence as the following,
\begin{equation}
 \frac{t(m)}{\rm yr} = 1.3\times10^{10}\left(\frac{m}{\MO}\right)
                                     \left(\frac{\LO}{l(m)}\right)\,.
\end{equation}
In this Letter, we choose the life time of star formation as the
luminosity-weighted average life time of OB main-sequence stars.
This is calculated by the following equation,
\begin{equation}
 \langle t \rangle = \frac{\int_{3\MO}^{100\MO}t(m)l(m)\psi(m)dm}
                          {\int_{3\MO}^{100\MO}l(m)\psi(m)dm}\,,
\end{equation}
then, we obtain $\langle t \rangle = 3.3 \times 10^6$ yr.

Finally, from equations (\ref{lob2}) and (\ref{stellarmass}), we derive
the following formula for the SFR of a galaxy in terms of its dust-IR
luminosity by adopting $3.3\times10^6{\rm yr}$ of the star formation
time-scale,
\begin{equation}
 \frac{{\rm SFR}}{\MO\, {\rm yr^{-1}}}=
  \frac{3.3\times 10^{-10}(1-\eta)}
       {0.4-0.2f+0.6\,\epsilon}\,
  \frac{L{\rm ^{obs}_{IR}}(8-1000\mu{\rm m})}{\LO}.       
 \label{sfr}
\end{equation}
This is the main result of this Letter.

\section{Discussions}

As stated in \S 1, the previous works on the SFR in IR luminosity is not
applicable to galaxies with a moderate star formation activity, because
we can no longer assume their IR luminosity to be their bolometric one
for such galaxies.
On the contrary, our equation (\ref{sfr}) is also reasonable for them.
As a demonstration of our equation (\ref{sfr}), hence, we apply it to
galaxies with a moderate star formation activity.
We can adopt the suitable sample with the data set of the IRAS
luminosity, $L{\rm^{obs}_{FIR}}(40-120\mu{\rm m})$, in Usui et
al. (1998), for example.
Their sample contains 15 early (Sa--Sab) spiral galaxies showing rather
high star forming activity in far-IR, and their averaged IRAS luminosity
is about $5.8\times10^9\LO$.

For absorbing fraction by neutral hydrogen, $f$, we adopt 0.26 derived by
Petrosian et al. (1972) for Orion nebula.
The efficiency of dust absorption for nonionizing photons,
$\epsilon$, is estimated to be 0.6 from the averaged 1000--4000\AA\
extinction curve of the Galaxy (Savage \& Mathis 1979) and the
average visual extinction of Usui's sample ($A_V = 1$ mag from private
communication with Usui).
Here, we should note that when our formula is applied to
the individual molecular clouds, these parameters may depend on the geometry 
of clouds.
Moreover, we choose 0.5 for the cirrus fraction, $\eta$, of Usui's sample,
according to a model of Lonsdale-Persson \& Helou (1987).
In addition, we have converted $L{\rm ^{obs}_{IR}}(8-1000\mu{\rm m})$ to 
$L{\rm ^{obs}_{FIR}}(40-120\mu{\rm m})$ calculated easily from IRAS 60
and 100 $\mu$m fluxes by the factor of 1.4, assuming the modified black body
radiation ($I_\nu \propto \nu B_\nu$, where $I_\nu$, $\nu$, and $B_\nu$
are intensity, frequency, and the Plank function, respectively) of 30 K
from an optically thin dust medium.
Of course, this factor depends on the dust temperature.
For example, if we assume 50 K and 15 K, the factors are changed to 1.1
and 5.6, respectively.
Thus, we must note this point when our formula is applied to galaxies
whose IR emission is dominated by almost only cold dust (e.g., 15 K)
owing to their quiescent star forming activity.

Accordingly we obtain 2 $\MO\,{\rm yr^{-1}}$ as the average value for
the sample galaxies in Usui et al. (1998).
This SFR is very similar to the value, 1.4 $\MO\,{\rm yr^{-1}}$, for the
same sample via the equation for starburst galaxies in Kennicutt
(1998b).
Thus, we find coincidence between the two estimated SFRs.
This coincidence originates from the following two reasons.
[1] The coefficient of conversion from IR luminosity to SFR tends to
decrease due to the cirrus fraction of IR. 
[2] It also tends to increase due to the lower optical depth of dust.
Thus, we suggest that since the effect of the cirrus
fraction is very effective, that of the lower optical depth of dust is
offset for adopted early spiral galaxies.

Now we will discuss what type of galaxies our result can be applied to.
Our algorithm starts from the expected dust-IR luminosity of an H II
region.
There, we assume that the dust content is enough to absorb
Lyman-$\alpha$ photons before they run away from this region.
For most observed H II region, the Case B approximation is applicable,
thus, all Lyman-$\alpha$ photons will be absorbed by dust within this
region, even with a relatively low amount of dust (Spitzer 1978).
Hence, equation (\ref{sfr}) can be applied to galaxies with star
formation in H II regions if they have neither an extremely small
dust-to-gas ratio nor an active galactic nucleus (AGN).
However, we can adapt equation (\ref{sfr}) to the AGN plus star
forming galaxies, subtracting the AGN component of dust-IR luminosity from
the observed IR luminosity.
On the other hand, the limit of dust-to-gas ratio for the application of
equation (\ref{sfr}) will be examined in our future work.

Finally, we discuss the choice of an IMF, and an upper and lower
cut-offs of mass.
We examine the effect of variance of the IMF, the upper limit
mass or the lower limit mass. 
When we change only the IMF for the Scalo's IMF obtained from
Binney \& Merrifield (1998), the coefficient of the equation (\ref{sfr})
is smaller than that of the Salpeter's IMF by the factor of 0.8.
If we substitute 60 $\MO$ into the upper limit mass, the recent SFR for 
the sample of Usui et al, (1998) becomes 1.4 times larger than
that of 100 $\MO$.
If the lower cut off mass is set 1 $\MO$, the coefficient of
the equation (\ref{sfr}) decrease by the factor of 0.4.
In summary, the choices of an IMF, an upper limit mass and a lower limit
mass cause uncertainties of the factor of about 2 to our result.

\section{Conclusions}

We summarize the conclusions reached in this Letter:

1. Starting from the dust-IR luminosity expected theoretically from a
dusty H II region with the Case B approximation, we
formulate equation (\ref{sfr}) to estimate the SFR of galaxies from their
observed IR luminosity.

2. This new simple formula for SFR contains three parameters explicitly:
[1] the fraction of ionizing photons from young massive stars absorbed by
neutral hydrogen in star forming regions, $f$, [2] the averaged
efficiency of dust absorption for nonionizing photons from young massive
stars, $\epsilon$,  and [3] the cirrus fraction of observed IR
luminosity, $\eta$.

3. Using equation (\ref{sfr}) and adopting a proper set of three
parameters, the recent SFR averaged for the sample in Usui et
al. (1998), which consists of 15 early (Sa--Sab) spiral galaxies with
moderate IR luminosity ($\sim10^{9-10}\LO$), is calculated to be about 2
$\MO\,{\rm yr^{-1}}$.

4. The derived convenient equation (\ref{sfr}) is applied to any
galaxies forming young massive stars in H II regions as far as they have
neither an extremely small dust-to-gas ratio nor an AGN, if we choose a
set of parameters reasonable for applied galaxies.

5. The coefficient of equation (\ref{sfr}) has uncertainty
of the factor of about 2 by the choices of a specific IMF, and its upper
and lower limits of stellar mass.

\vspace{1pc}\par
We would like to thank Prof.\ M.\ Sait\={o} for
continuous encouragement and Dr.\ T.\ Usui for informative discussions.
One of us (H.H.) acknowledges the Research Fellowship of
the Japan Society for the Promotion of Science for Young
Scientists.

\section*{References}
\small

\re
 Binney J., Merrifield M.\
 1998, Galactic Astronomy (New Jersey: Princeton) Chap.3, 5

\re
 Buat V., Xu C.\ 1996, A\&A 306, 61

%\re
% de Vaucouleurs G., de Vaucouleurs A., Corwin H.G., Buta R.J., Patural
% G., Fouqe\'{e} P.\ 
% 1991, Third Reference Catalogue of Bright Galaxies (New York: Springer)

\re
 Devereux N.A., Young J.S.\ 1991, ApJ 371, 515

\re
 Efstathiou A., Rowan-Robinson M., Siebenmorgen R.\ 2000, MNRAS 313, 734

\re
 Helou G.\ 1986, ApJ 311, L33

\re
 Kennicutt R.C.\ 1998a, ARA\&A 36, 189

\re
 Kennicutt R.C.\ 1998b, ApJ 498, 541

\re
 Lonsdale-Persson C.J., Helou G.\ 1987, ApJ 314, 513

\re
 Osterbrock D.E.\ 1989, Astrophysics of Gaseous Nebulae and Active
 Galactic Nuclei (Mill Valley: University Science Books) Chap.4

\re
 Panagia N.\ 1973, AJ 78, 929

\re
 Petrosian V., Silk J., Field G.B.\ 1972, ApJ 177, L69

\re
 Salpeter E.E.\ 1955, ApJ 121, 161

\re
 Savage B.D., Mathis J.S.\ 1979, ARA\&A 17, 73

\re
 Schwarzchild M.\ 1958, Structure and Evolution of the Stars (New York: 
 Dover)

\re
 Silva L., Granato G.L., Bressan A., Danese L.\ 1998, ApJ 509, 103

\re
 Soifer B.T., Houck J.R., Neugebauer G.\ 1987, ARA\&A 25, 187
 
\re
 Spitzer L.\ 
 1978, Physical Processes in the Interstellar Medium (New York: Wiley)
 Chap.5, 7

\re
 Thronson H.A., Telesco C.M.\ 1986, ApJ 311, 98

\re
 Usui T., Sait\={o} M., Tomita A.\ 1998, AJ 116, 2166

\end{document}